\documentclass[aps,12pt,pra,showpacs,superscriptaddress,preprint]{revtex4}

\usepackage{amsmath}

\usepackage[dvipdfmx]{graphicx}

\usepackage{subfig}

\usepackage{array}

\usepackage{amsfonts}

\usepackage{epsf}

\usepackage{epsfig}

\usepackage{amssymb}

\usepackage{bbm}

\usepackage{delarray}

\usepackage{caption}

\usepackage{amstext}

\usepackage{graphics}

\usepackage{epstopdf}

\begin{document}

\title{A General Approach for the Exact Solution of the Schrodinger Equation}
\author{\small Cevdet Tezcan}
\affiliation{Faculty of Engineering, Baskent University, Baglica Campus, Ankara,Turkey}
\author{\small Ramazan Sever}
\email[E-mail: ]{sever@metu.edu.tr}\affiliation{Department of
Physics, Middle East Technical  University, 06531, Ankara,Turkey}

\begin{abstract}
The Schrodinger equation is solved exactly for some well known
potentials. Solutions are obtained reducing the Schrodinger
equation into a second order differential equation by using an
appropriate coordinate transformation. The Nikiforov-Uvarov method
is used in the calculations to get energy eigenvalues and the
corresponding wave
functions.
\end{abstract}


\maketitle

\section{Introduction}
The Schrodinger equation(SE) is one of the fundamental wave
equations in physics. Its solutions for some certain potentials have
important applications in atomic, nuclear, condensed and high energy
physics and particle physics[1-26].

Several methods are used in the solution of the Schrodinger
equation. One of them is the Nikiforov-Uvarov(NU) method[27]. It
provides us an exact solution of eigenvalues and eigenfunctions.

In this work, we get parametric solution of the SE equation by
transforming into a second order parametric differential equation
having a certain form. This approach is independent of the potential
solved. In the procedure, There is no need to calculate the main
parameters of the NU method for each solution of the potential.

We solve the eight well known potentials to calculate energy
eigenvalues and the corresponding wave functions exactly. These
potentials are Morse[28], Rosen-Morse[29], Pseudoharmonic[30],
Mie[31-34 ], Woods-Saxon[35], Poschl-Teller[36, 37],
Kratzer-Fues[38], Noncentral[39, 40]. Woods-Saxon potential
describes the interaction of a neutron with a heavy nucleus. The
noncentral potential is used to describe bound states of an electron
in the Coulomb field together with Aharonov-Bohm field and/or Dirac
monopole. Other potentials are mainly used to describe bound states
of spectroscopy of die-atomic molecules.

The contents of the present paper is as follows: In section II,
Nikiforov-Uvarov method is summarized. In section III, Solutions are
presented. Results are discussed in section IV.

\section{Nikiforov-Uvarov Method}

The Schrodinger equation is transformed into a second order
differential equation of the form with an appropriate coordinate
transformation of the form

\begin{equation} \label{eq.2}
\frac{d^{2}\psi(s)}{ds^{2}}+\frac{\tilde{\tau}(s)}{\sigma(s)}
\frac{d\psi(s)}{ds}+\frac{\tilde{\sigma}(s)}{{\sigma}^{2}(s)}\psi(s)=0
\end{equation}
where $\sigma(s)$, $\tilde{\sigma}(s)$ are polynomials at most
second degree and $\tilde{\tau}(s)$ is a first degree polynomial. In
this method, one defines

\begin{equation} \label{eq.3}
\pi(s)=\frac {\left(\sigma^{\prime}-\tilde{\tau}\right)}{2}\pm
\sqrt{\left(\frac{\sigma^{\prime}-\tilde{\tau}}{2}\right)^{2}-
\tilde{\sigma}+k\sigma},
\end{equation}
and

\begin{equation} \label{eq.4}
\lambda=k+\pi^{\prime}(s ),
\end{equation}
where $\lambda$ and $k$ are constants. Since square root in the the
polynomial $\pi$ in Eq. (2) must be a square then this defines the
constant $k.$ Replacing $k$ into Eq. (2), we define

\begin{equation} \label{eq.5}
\tau(s)=\tilde{\tau}(s)+2\pi(s).
\end{equation}
Since $\rho(s)>0$ and $\sigma(s)>0$, hence the derivative of
$\tau$ should be negative[27]. This leads the choice of the
solution. If $\lambda$ in Eq. (3) is

\begin{equation} \label{eq.6}
\lambda=\lambda_{n}=-n\tau^{\prime}-\frac{\left[n(n-1)\sigma^{\prime\prime}\right]}{2},
\quad n=0,1,2,\ldots
\end{equation}
The hypergeometric type equation has a particular solution with
degree $n$. Solution of Eq. (1) can be obtained with the product of
two independent parts

\begin{equation} \label{eq.7}
\psi(s)=\phi(s)~y(s),
\end{equation}
where $y(s)$ can be written as

\begin{equation} \label{eq.8}
y_{n}(s)=\frac{B_{n}}{\rho(s)}\frac{d^{n}}{ds^{n}}
\left[\sigma^{n}(s)~\rho(s)\right],
\end{equation}
and $\rho(s)$ should satisfy the condition

\begin{equation} \label{eq.9}
\frac{d}{ds}\left[\sigma(s)~\rho(s)\right]=\tau(s)~\rho(s).
\end{equation}
The other factor is defined as

\begin{equation} \label{eq.10}
\frac{\phi^{\prime}(s)}{\phi(s)}=\frac{\pi(s)}{\sigma(s)}.
\end{equation}
The following equation is a general form of the Schr\&quot;{o}dinger
equation written for any potentials

\begin{equation} \label{eq.11}
\left[\frac{d^{2}}{ds^{2}}+\frac{\alpha_{1}-\alpha_{2}s}{s(1-\alpha_{3}s)}\frac{d}{ds}+\frac{-\xi_{1}s^{2}+\xi_{2}s-\xi_{3}}{[s(1-\alpha_{3}s)]^{2}}\right]\psi=0.
\end{equation}
We may solve this as follows. When Eq. (10) is compared with Eq.
(1), we get

\begin{equation} \label{eq.14}
\tilde{\tau}=\alpha_{1}-\alpha_{1}s
\end{equation}
and

\begin{equation} \label{eq.15}
\sigma=s(1-\alpha_{3}s)
\end{equation}
also

\begin{equation} \label{eq.16}
\tilde{\sigma}=-\xi_{1}s^{2}+\xi_{2}s-\xi_{3}.
\end{equation}
Substituting these into Eq. (2)

\begin{equation} \label{eq.17}
\pi=\alpha_{4}+\alpha_{5}s\pm\sqrt{(\alpha_{6}-k\alpha_{3})s^{2}+(\alpha_{7}+k)s+\alpha_{8}}
\end{equation}
where

\begin{equation} \label{eq.18}
\alpha_{4}=\frac{1}{2}(1-\alpha_{1}),
\end{equation}

\begin{equation} \label{eq.19}
\alpha_{5}=\frac{1}{2}(\alpha_{2}-2\alpha_{3}),
\end{equation}

\begin{equation} \label{eq.20}
\alpha_{6}=\alpha_{5}^{2}+\xi_{1},
\end{equation}

\begin{equation} \label{eq.21}
\alpha_{7}=2\alpha_{4}\alpha_{5}-\xi_{2},
\end{equation}

\begin{equation} \label{eq.22}
\alpha_{8}=\alpha_{4}^{2}+\xi_{3}.
\end{equation}
In Eq. (14), the function under square root must be the square of a
polynomial according to the NU method, so that

\begin{equation} \label{eq.23}
k_{1,2}=-(\alpha_{7}+2\alpha_{3}\alpha_{8})\pm2\sqrt{\alpha_{8}\alpha_{9}},
\end{equation}
Where, we define

\begin{equation} \label{eq.24}
\alpha_{9}=\alpha_{3}\alpha_{7}+\alpha_{3}^{2}\alpha_{8}+\alpha_{6}.
\end{equation}
For each $k$ the following $\pi$'s are obtained. For

\begin{equation} \label{eq.25}
k=-(\alpha_{7}+2\alpha_{3}\alpha_{8})-2\sqrt{\alpha_{8}\alpha_{9}}
\end{equation}
$\pi$ becomes:

\begin{equation} \label{eq.26}
\pi=\alpha_{4}+\alpha_{5}s-\left[(\sqrt{\alpha_{9}}+\alpha_{3}\sqrt{\alpha_{8}})s-\sqrt{\alpha_{8}}\right].
\end{equation}
For the same $k$, from Eqs.(4), (11) and (14)

\begin{equation} \label{eq.27}
\tau=\alpha_{1}+2\alpha_{4}-(\alpha_{2}-2\alpha_{5})s-2\left[(\sqrt{\alpha_{9}}+\alpha_{3}\sqrt{\alpha_{8}})s-\sqrt{\alpha_{8}}\right]
\end{equation}
and

\begin{eqnarray} \label{eq.28}
\tau^{\prime}&=&-(\alpha_{2}-2\alpha_{5})-2(\sqrt{\alpha_{9}}+\alpha_{3}\sqrt{\alpha_{8}})
\nonumber \\
                              &=&-2\alpha_{3}-2(\sqrt{\alpha_{9}}+\alpha_{3}\sqrt{\alpha_{8}})
\end{eqnarray}
are obtained. When Eq.(3) is used with Eqs. (24) and (25) the
following equation is derived:

\begin{eqnarray} \label{eq.29}
\alpha_{2}n-(2n+1)\alpha_{5}&+&(2n+1)(\sqrt{\alpha_{9}}+\alpha_{3}\sqrt{\alpha_{8}})+n(n-1)\alpha_{3}\nonumber
\\
&+&\alpha_{7}+2\alpha_{3}\alpha_{8}+2\sqrt{\alpha_{8}\alpha_{9}}=0.
\end{eqnarray}
This equation gives the energy spectrum of a given problem. From Eq.
(8)

\begin{equation} \label{eq.30}
\rho(s)=s^{\alpha_{10}-1}(1-\alpha_{3}s)^{\frac{\alpha_{11}}{\alpha_{3}}-\alpha_{10}-1}
\end{equation}
is found and when this equation is used in Eq. (7)

\begin{equation} \label{eq.31}
y_{n}=P_{n}^{(\alpha_{10}-1,\frac{\alpha_{11}}{\alpha_{3}}-\alpha_{10}-1)}(1-2\alpha_{3}s)
\end{equation}
is obtained, where,

\begin{equation} \label{eq.32}
\alpha_{10}=\alpha_{1}+2\alpha_{4}+2\sqrt{\alpha_{8}}
\end{equation}
and

\begin{equation} \label{eq.33}
\alpha_{11}=\alpha_{2}-2\alpha_{5}+2(\sqrt{\alpha_{9}}+\alpha_{3}\sqrt{\alpha_{8}})
\end{equation}
and $P_{n}^{(\alpha,\beta)}$ are Jacobi polynomials. Using Eq.(9)

\begin{equation} \label{eq.34}
\phi(s)=s^{\alpha_{12}}(1-\alpha_{3}s)^{-\alpha_{12}-\frac{\alpha_{13}}{\alpha_{3}}}
\end{equation}
is obtained and the general solution becomes:

\begin{equation} \label{eq.35}
\psi=\phi(s)y(s)
\end{equation}

\begin{equation} \label{eq.36}
\psi=s^{\alpha_{12}}(1-\alpha_{3}s)^{-\alpha_{12}-\frac{\alpha_{13}}{\alpha_{3}}}P_{n}^{(\alpha_{10}-1,\frac{\alpha_{11}}{\alpha_{3}}-\alpha_{10}-1)}(1-2\alpha_{3}s).
\end{equation}
Here, alpha functions are given by:

\begin{equation} \label{eq.37}
\alpha_{12}=\alpha_{4}+\sqrt{\alpha_{8}}
\end{equation}
and

\begin{equation} \label{eq.38}
\alpha_{13}=\alpha_{5}-(\sqrt{\alpha_{9}}+\alpha_{3}\sqrt{\alpha_{8}})
\end{equation}
 In some problems $\alpha_{3}=0.$ For this type of problems when

\begin{equation} \label{eq.39}
\lim_{\alpha_{3} \to
0}P_{n}^{(\alpha_{10}-1,\frac{\alpha_{11}}{\alpha_{3}}-\alpha_{10}-1)}(1-\alpha_{3}s)=L_{n}^{\alpha_{10}-1}(\alpha_{11}s)
\end{equation}
and

\begin{equation} \label{eq.40}
\lim_{\alpha_{3} \to
0}(1-\alpha_{3}s)^{-\alpha_{12}-\frac{\alpha_{13}}{\alpha_{3}}}=e^{\alpha_{13}s},
\end{equation}
the solution given in Eq. (33) becomes as

\begin{equation} \label{eq.41}
\psi=s^{\alpha_{12}}e^{\alpha_{13}s}L_{n}^{\alpha_{10}-1}(\alpha_{11}s).
\end{equation}

In some cases, one may need a second solution of Eq. (10). In this
case, if the same procedure is followed by using

\begin{equation}
k=-(\alpha_{7}+2\alpha_{3}\alpha_{8})+2\sqrt{\alpha_{8}\alpha_{9}}
\end{equation}
the solution becomes

\begin{equation}
\psi=s^{\alpha_{12}^{*}}(1-\alpha_{3}s)^{-\alpha_{12}^{*}-\frac{\alpha_{13}^{*}}{\alpha_{3}}}P_{n}^{(\alpha_{10}^{*}-1,\frac{\alpha_{11}^{*}}{\alpha_{3}}-\alpha_{10}-1)}(1-2\alpha_{3}s)
\end{equation}
and the energy spectrum is

\begin{equation}
\alpha_{2}n-2\alpha_{5}n+(2n+1)(\sqrt{\alpha_{9}}-\alpha_{3}\sqrt{\alpha_{8}})+n(n-1)\alpha_{3}+\alpha_{7}+2\alpha_{3}\alpha_{8}-2\sqrt{\alpha_{8}\alpha_{9}}-\alpha_{5}=0
\end{equation}
Pre-defined $\alpha$ parameters are:

\begin{equation} \label{eq.66}
\begin{array}{lcr}
\alpha_{10}^{*}=\alpha_{1}+2\alpha_{4}-2\sqrt{\alpha_{8}}, &&  \\
\alpha_{11}^{*}=\alpha_{2}-2\alpha_{5}+2(\sqrt{\alpha_{9}}-\alpha_{3}\sqrt{\alpha_{8}}),
&& \\
\alpha_{12}^{*}=\alpha_{4}-\sqrt{\alpha_{8}}, &&  \\
\alpha_{13}^{*}=\alpha_{5}-(\sqrt{\alpha_{9}}-\alpha_{3}\sqrt{\alpha_{8}})
&&
\end{array}
\end{equation}

\section{Some Applications}

Case 1:  Generalized Morse Potential

We use the Generalized Morse Potential[28]

\begin{equation} \label{eq.42}
V(x)=V_{1}e^{-2\alpha x}-V_{2}e^{-\alpha x}
\end{equation}
for the transformation $s=\sqrt{V_{1}}e^{-\alpha x}$, the
Schrodinger equation becomes

\begin{equation} \label{eq.43}
\frac{d^{2}\psi}{ds^{2}}+\frac{1}{s}\frac{d\psi}{ds}-\left(\frac{2m}{\hbar^{2}\alpha^{2}}s^{2}-\frac{2m}{\hbar^{2}\alpha^{2}}\frac{V_{2}}{\sqrt{V_{1}}}s+4\varepsilon^{2}\right)\frac{\psi}{s^2}=0
\end{equation}

where
\begin{equation} \label{eq.44'}
\varepsilon^{2}=-\frac{mE}{2\hbar^{2}\alpha^{2}}
\end{equation}
When Eq. (40) is compared with Eq.(10), we get

\begin{equation} \label{eq.45}
\begin{array}{lccr}
\alpha_{1}=1, &; \alpha_{2}=0, &; \alpha_{3}=0, &; \alpha_{4}=0 \\
\alpha_{5}=0, &; \alpha_{6}=\xi_{1}, &; \alpha_{7}=-\xi_{2}, &; \alpha_{8}=\xi_{3}  \\
\alpha_{9}=\xi_{1}, &; \alpha_{10}=1+2\sqrt{\xi_{3}}, &; \alpha_{11}=2\sqrt{\xi_{1}}  \\
\alpha_{12}=\sqrt{\xi_{3}}, &; \alpha_{13}=-\sqrt{\xi_{1}}
\end{array}.
\end{equation}
and

\begin{equation} \label{eq.45}
\xi_{1}=\frac{2m}{\hbar^{2}\alpha^{2}}, \quad
\xi_{2}=\frac{2m}{\hbar^{2}\alpha^{2}}\frac{V_{2}}{\sqrt{V_{1}}},\quad
\xi_{3}=4\varepsilon^{2}
\end{equation}
Using Eqs. (46,47), we calculate energy eigenvalues and the
corresponding wave functions as ($\hbar^2/(2m)=1$)

From Eqs. (26) and (38), we obtain the parameters
$(\alpha_{1}-\alpha_{13})$ and $(\xi_{1}-\xi_{3})$ so energy
eigenvalues become

\begin{equation} \label{eq.47}
E=-\frac{1}{4}\alpha^{2} \left[2n+1-\frac{V_{2}}{\alpha\sqrt{V_{1}}}\right]
\end{equation}
and the corresponding wave functions take

\begin{equation} \label{eq.48}
\psi=s^{2\varepsilon}e^{-\frac{1}{\alpha}s}L_{n}^{4\varepsilon}(2\gamma s)
\end{equation}

Case 2: Deformed Rosen-Morse Potential

Deformed Rosen-Morse potential[29] has the form

\begin{equation} \label{eq.48}
V(x)=\frac{V_{1}}{1+qe^{-2\alpha x}}-V_{2}q \frac{e^{-2\alpha
x}}{(1+qe^{-2\alpha x})^{2}}.
\end{equation}
The SE becomes

\begin{equation} \label{eq.65}
\frac{d^{2}\psi}{ds^{2}}+\frac{1-qs}{s(1-qs)}\frac{d\psi}{ds}+\frac{1}{[s(1-qs)]^{2}}\left[-\varepsilon
q^{2}s^{2}+(2\varepsilon q+\beta q-\gamma)s-(\varepsilon+\beta)\right]\psi=0
\end{equation}
We define the parameters as

\begin{equation} \label{eq.66}
\begin{array}{lcr}
\xi_{1}=\varepsilon q^{2}, &; \xi_{2}=2\varepsilon q+\beta q-\gamma, &
\xi_{3}=\varepsilon+\beta \\
\alpha_{1}=1, &; \alpha_{2}=q, &; \alpha_{3}=q \\
\alpha_{4}=0, &; \alpha_{5}=-\frac{q}{2}, &\alpha_{6}=\frac{q^{2}}{4}+\xi_{1} \\
\alpha_{7}=-\xi_{2}, &; \alpha_{8}=\xi_{3},\\
\alpha_{9}=\xi_{1}-q\xi_{2}+q^{2}\xi_{3}+\frac{q^{2}}{4}, &\\
\alpha_{10}=1+2\sqrt{\xi_{3}}, \\
\alpha_{11}=2q+2(\sqrt{\xi_{1}-q\xi_{2}+q^{2}\xi_{3}+\frac{q^{2}}{4}}+q\sqrt{\xi_{3}}),
\\
\alpha_{12}=\sqrt{\xi_{3}}, \\
\alpha_{13}=-\frac{q}{2}-(\sqrt{\xi_{1}-q\xi_{2}+q^{2}\xi_{3}+\frac{q^{2}}{4}}+q\sqrt{\xi_{3}})
\end{array}
\end{equation}
The energy spectrum is

\begin{equation} \label{eq.65}
\varepsilon=-\frac{\beta}{2}+\frac{1}{16}\left(2n+1+\sqrt{1+\frac{4\gamma}{q}}\right)^{2}+\left(\frac{\beta}{2n+1+\sqrt{1+\frac{4\gamma}{q}}}\right)^{2}
\end{equation}
and the wave function becomes

\begin{equation} \label{eq.63}
\psi=s^{\sqrt{\varepsilon+\beta}}(1-qs)^{\frac{1}{2}\left[1+\sqrt{1+\frac{4\gamma}{q^{2}}}\right]}P_{n}^{(2\sqrt{\varepsilon+\beta},\sqrt{1+\frac{4\gamma}{q}})}(1-2qs)
\end{equation}

Case 3:  Pseudoharmonic Potential

For pseudoharmonic Potential is[30]

\begin{equation} \label{eq.50}
V(r)=V_{0} \left(\frac{r}{r_{0}}-\frac{r_{0}}{r}\right)^{2}.
\end{equation}
By using the transformation $s=r^{2}$, the radial part of the
Schrodinger equation becomes:

\begin{equation} \label{eq.51}
\frac{d^{2}R}{ds^{2}}+\frac{3/2}{s}\frac{dR}{ds}+\frac{-\alpha^{2}s^{2}+\varepsilon
s-\beta}{s^{2}}R(s)=0
\end{equation}
By using the following parameters

\begin{equation} \label{eq.52}
\begin{array}{lcr}
\xi_{1}=\alpha^{2}, &; \xi_{2}=\varepsilon, &; \xi_{3}=\beta \\
\alpha_{1}=\frac{3}{2}, &; \alpha_{2}=0, &; \alpha_{3}=0 \\
\alpha_{4}=-\frac{1}{4}, &; \alpha_{5}=0, &; \alpha_{6}=\xi_{1} \\
\alpha_{7}=-\xi_{2}, &; \alpha_{8}=\frac{1}{16}+\xi_{3}, &; \alpha_{9}=\xi_{1} \\
\alpha_{10}=1+2\sqrt{\frac{1}{16}+\xi_{3}}, &; \alpha_{11}=2\alpha, &;
\alpha_{12}=-\frac{1}{4}+\sqrt{\frac{1}{16}+\xi_{3}} \\
\alpha_{13}=-\sqrt{\xi_{1}} \end{array}
\end{equation}

\begin{equation} \label{eq.53}
\varepsilon=\left[2n+1+2\sqrt{\beta+\frac{1}{16}}\right]\alpha
\end{equation}
and

\begin{equation} \label{eq.54}
\psi=s^{-\frac{1}{4}+\sqrt{\frac{1}{16}+\beta}} e^{-\alpha
s}L_{n}^{2\sqrt{\frac{1}{16}+\beta}} (2\alpha s)
\end{equation}
are obtained.

Case 4: Mie Potential

Mie Potential has the form [31-34]

\begin{equation} \label{eq.55}
V(r)=V_{0}
\left[\frac{1}{2}\left(\frac{a}{r}\right)^{2}-\frac{a}{r}\right].
\end{equation}
By using the transformation $r=s$, we obtain the radial part of the
Schrodinger equation as:

\begin{equation} \label{eq.56}
\frac{d^{2}R}{ds^{2}}+\frac{2}{s}\frac{dR}{ds}+\frac{\varepsilon^{2}s^{2}-\beta
s-\gamma}{s^{2}}R=0
\end{equation}
where

\begin{equation} \label{eq.57}
\beta=-\frac{2\mu}{\hbar^{2}},\quad
\gamma=\frac{2\mu}{\hbar^{2}}\left[\frac{1}{2}V_{0}a^{2}+\frac{l(l+1)\hbar^{2}}{2\mu}\right].
\end{equation}
By using the parameters

\begin{equation} \label{eq.58}
\begin{array}{lcr}
\xi_{1}=-{\epsilon}^{2}, &; \xi_{2}=-\beta, &; \xi_{3}=\gamma \\
\alpha_{1}=2, &; \alpha_{2}=0, &; \alpha_{3}=0 \\
\alpha_{4}=-\frac{1}{2}, &; \alpha_{5}=0, &; \alpha_{6}=\xi_{1} \\
\alpha_{7}=-\xi_{2}, &; \alpha_{8}=\frac{1}{4}+\xi_{3}, &; \alpha_{9}=\xi_{1} \\
\alpha_{10}=1+2\sqrt{\frac{1}{4}+\xi_{3}}, &; \alpha_{11}=2\sqrt{\xi_{1}}, &;
\alpha_{12}=-\frac{1}{2}+\sqrt{\frac{1}{4}+\xi_{3}} \\
\alpha_{13}=-\sqrt{\xi_{1}} \end{array}
\end{equation}
We get energy eigenvalues and corresponding wave functions as

\begin{equation} \label{eq.59}
-\varepsilon^{2}=\beta^{2}\left[2n+1+\sqrt{1+4\gamma}\right]^{-2}
\end{equation}
and

\begin{equation} \label{eq.60}
\psi=As^{-\frac{1}{2}(1-\sqrt{1+4\gamma})}e^{-i\varepsilon
s}L_{n}^{\sqrt{1+4\gamma}}(2i\varepsilon s)
\end{equation}

Case 5: Generalized Woods-Saxon Potential

The generalized Woods-Saxon potential has the form [35]

\begin{equation}
V(r)=-\frac{V_{0}}{1+e^{\left(\frac{r-R_{0}}{a}\right)}}-\frac{C.e^{\left(\frac{r-R_{0}}{a}\right)}}{\left(1+e^{\left(\frac{r-R_{0}}{a}\right)}\right)^2}~,
\end{equation}

The SE takes the following form, after applying a coordinate
transformation
$s=\frac{1}{1+e^{\left(\frac{r-R_{0}}{a}\right)}}$,

\begin{equation} \label{eq.65}
\frac{d^{2}\psi}{ds^{2}}+\frac{1-2s}{s(1-s)}\frac{d\psi}{ds}+\frac{1}{[s(1-s)]^{2}}\left[-\varepsilon+\beta
s+\gamma s(1-s)\right]\psi=0
\end{equation}
where

\begin{equation}
\varepsilon=-\frac{2ma^{2}}{\hbar^{2}}E>0, \quad
\beta=\frac{2ma^{2}V_{0}}{\hbar^{2}}, \quad \gamma=\frac{2ma^{2}}{\hbar^{2}}C
\end{equation}
By using the following parameter values

\begin{equation} \label{eq.61}
\begin{array}{lcr}
\xi_{1}=\gamma, &; \xi_{2}=\beta+\gamma, &; \xi_{3}=\varepsilon \\
\alpha_{1}=1, &; \alpha_{2}=2, &; \alpha_{3}=1 \\
\alpha_{4}=0, &; \alpha_{5}=0, &; \alpha_{6}=\xi_{1} \\
\alpha_{7}=-\xi_{2}, &; \alpha_{8}=\xi_{3}, &; \alpha_{9}=\xi_{1}-\xi_{2}+\xi_{3} \\
\alpha_{10}=1+2\sqrt{\xi_{3}}, &;
\alpha_{11}=2+2(\sqrt{\xi_{1}-\xi_{2}+\xi_{3}}+\sqrt{\xi_{3}}), &;
\alpha_{12}=\sqrt{\xi_{3}} \\
\alpha_{13}=-(\sqrt{\xi_{1}-\xi_{2}+\xi_{3}}+\sqrt{\xi_{3}}) \end{array}
\end{equation}
energy eigenvalues and corresponding wave functions are obtained as

\begin{equation} \label{eq.62}
\varepsilon=\frac{\beta^{2}}{\left[-(2n+1)+\sqrt{1+4\gamma}\right]^{2}}+\frac{\beta}{2}+\frac{1}{16}\left[-(2n+1)+\sqrt{1+4\gamma}\right]^{2}
\end{equation}

\begin{equation} \label{eq.63}
\psi=s^{\sqrt{\varepsilon}}(1-s)^{\sqrt{\varepsilon-\beta}}P_{n}^{(2\sqrt{\varepsilon},2\sqrt{\varepsilon-\beta})}(1-2s).
\end{equation}

Case 6: Poschl-Teller Potential

Poschl-Teller Potential is[36, 37]

\begin{equation} \label{eq.64}
V(x)=-4V_{0}\frac{e^{-2\alpha x}}{(1+q e^{-2\alpha x})^{2}}
\end{equation}
The SE equation has the form

\begin{equation} \label{eq.65}
\frac{d^{2}\psi}{ds^{2}}+\frac{1-qs}{s(1-qs)}\frac{d\psi}{ds}+\frac{1}{s^{2}(1-qs)^{2}}\left[-\varepsilon^{2}q^{2}s^{2}+(2\varepsilon^{2}q-\beta^{2})s-{\epsilon}^{2}\right]\psi=0
\end{equation}
The parameters take

\begin{equation} \label{eq.66}
\begin{array}{lcr}
\xi_{1}=\varepsilon^{2}q^{2}, &; \xi_{2}=2\varepsilon^{2}q-\beta^{2}, &;
\xi_{3}=\varepsilon^{2} \\
\alpha_{1}=1, &; \alpha_{2}=q, &; \alpha_{3}=q \\
\alpha_{4}=0, &; \alpha_{5}=-\frac{q}{2}, &; \alpha_{6}=\frac{q^{2}}{4}+\xi_{1} \\
\alpha_{7}=-\xi_{2}, &; \alpha_{8}=\xi_{3}, &; \\
\alpha_{9}=\xi_{1}-q\xi_{2}+q^{2}\xi_{3}+\frac{q^{2}}{4}, \\
\alpha_{10}=1+2\sqrt{\xi_{3}}, \\
\alpha_{11}=2q+2(\sqrt{\xi_{1}-q\xi_{2}+q^{2}\xi_{3}+\frac{q^{2}}{4}}+q\sqrt{\xi_{3}}),
\\
\alpha_{12}=\sqrt{\xi_{3}},\\
\alpha_{13}=-\frac{q}{2}-(\sqrt{\xi_{1}-q\xi_{2}+q^{2}\xi_{3}+\frac{q^{2}}{4}}+q\sqrt{\xi_{3}})
 \end{array}
\end{equation}
We get energy eigenvalues and the corresponding wave functions as

\begin{equation} \label{eq.67}
\varepsilon=-\frac{1}{4}\left[2n+1+\sqrt{1+\frac{4\beta^{2}}{q}}\right]
\end{equation}
also

\begin{equation} \label{eq.68}
E=-\frac{2\hbar^{2}\alpha^{2}}{m}\varepsilon^{2}
\end{equation}
and the the wave functions take the form

\begin{equation} \label{eq.69}
\psi=s^{\varepsilon}(1-2qs)^{\frac{1}{2}\left[1+\sqrt{1+\frac{4\beta^{2}}{q}}\right]}P_{n}^{(2\varepsilon,\sqrt{1+\frac{4\beta^{2}}{q}})}(1-2s)
\end{equation}

Case 7: Kratzer-Fues Potential

Kratzer-Fues potential is [38]

\begin{equation} \label{eq.79}
V(r)=D_{e}\left(\frac{r-r_{e}}{r}\right)^{2}.
\end{equation}
The SE has the form

\begin{equation} \label{eq.80}
\frac{d^{2}R_{nl}(r)}{dr^{2}}+\frac{2}{r}\frac{d
R_{nl}(r)}{dr}+\frac{1}{r^{2}}\frac{2\mu}{\hbar^{2}}\left[(E_{nl}-D_{e})r^{2}+2D_{e}r_{e}r
-(D_{e}r_{e}^{2}+\frac{l(l+1)\hbar^{2}}{2\mu})\right]R_{nl}(r)=0
\end{equation}
By defining the new parameters

\begin{equation} \label{eq.81}
\varepsilon^{2}=\frac{2\mu(E_{nl}-D_{e})}{\hbar^{2}}
\end{equation}

\begin{equation} \label{eq.82}
-\beta=\frac{4\mu D_{e}r_{e}}{\hbar^{2}}
\end{equation}

\begin{equation} \label{eq.83}
\gamma=\frac{2\mu
(D_{e}r_{e}^{2}+\frac{l(l+1)\hbar^{2}}{2\mu})}{\hbar^{2}}
\end{equation}
and by using

\begin{equation} \label{eq.84}
r=s
\end{equation}
The SE takes the form

\begin{equation} \label{eq.85}
\frac{d^{2}R_{nl}(s)}{ds^{2}}+\frac{2}{s}\frac{d
R_{nl}(s)}{ds}+\frac{1}{s^{2}}(\varepsilon^{2}s^{2}-\beta
s-\gamma)R_{nl}(s)=0.
\end{equation}
By defining the following parameters

 \begin{equation} \label{eq.86}
\begin{array}{lcr}
\xi_{1}=-\varepsilon^{2}, &; \xi_{2}=-\beta, &; \xi_{3}=\gamma \\
\alpha_{1}=2, &; \alpha_{2}=0, &; \alpha_{3}=0 \\
\alpha_{4}=-\frac{1}{2}, &; \alpha_{5}=0, &; \alpha_{6}=-\varepsilon^{2} \\
\alpha_{7}=\beta, &; \alpha_{8}=\frac{1}{4}+\gamma, &; \alpha_{9}=-\varepsilon^{2} \\
\alpha_{10}=1+2\sqrt{\gamma+\frac{1}{4}}, &; \alpha_{11}=2\sqrt{-\varepsilon^{2}}, &
\alpha_{12}=-\frac{1}{2}+\sqrt{\gamma+\frac{1}{4}} \\
\alpha_{13}=-\sqrt{-\varepsilon^{2}}. \end{array}
\end{equation}
We get energy eigenvalues and the corresponding wave functions as

\begin{equation} \label{eq.87}
-\varepsilon^{2}=\beta^{2}(2n+1+\sqrt{1+4\gamma})^{-2}
\end{equation}
and

\begin{equation} \label{eq.88}
R=A_{n}\nu^{-\frac{1}{2}(1-\eta)}e^{-\frac{\nu}{2}}L_{n}^{\eta}(\nu)
\end{equation}
where

\begin{equation} \label{eq.89}
\eta=\sqrt{1+4\gamma}), \quad \nu=2i\varepsilon s
\end{equation}

Case 8: Noncentral Potential

The noncentral potential is[39, 40]

\begin{equation} \label{eq.70}
V(r,\theta)=\frac{\alpha}{r}+\frac{\beta}{r^{2}sin^{2}\theta}+\gamma\frac{cos\theta}{r^{2}sin^{2}\theta}.
\end{equation}
It has also some special forms[41-46]. The radial part of the
Schr\&quot;{o}dinger equation becomes:

\begin{equation} \label{eq.71}
\frac{d^{2}R(r)}{dr^{2}}+\frac{2}{r}\frac{dR(r)}{dr}+\frac{2mr^{2}}{\hbar^{2}}\left[E-\frac{\alpha}{r}-\frac{\lambda}{r^{2}}\right]R(r)=0
\end{equation}
if

\begin{equation} \label{eq.72}
r=s
\end{equation}
transformation is used with

\begin{equation} \label{eq.73}
R_{n,l}(r)=\frac{1}{r}F_{n,l}(r).
\end{equation}
Thus, it becomes

\begin{equation} \label{eq.74}
\left[\frac{d^{2}}{ds^{2}}+\frac{1}{s^{2}}(-\varepsilon^{2}s^{2}-b^{2}s-\lambda)\right]F=0.
\end{equation}
By using the parameters

\begin{equation} \label{eq.75}
\begin{array}{lcr}
\xi_{1}=\varepsilon^{2}, &; \xi_{2}=-b^{2}, &; \xi_{3}=\lambda \\
\alpha_{1}=0, &; \alpha_{2}=0, &; \alpha_{3}=0 \\
\alpha_{4}=\frac{1}{2}, &; \alpha_{5}=0, &; \alpha_{6}=\xi_{1} \\
\alpha_{7}=-\xi_{2}, &; \alpha_{8}=\frac{1}{4}+\xi_{3}, &; \alpha_{9}=\xi_{1} \\
\alpha_{10}=1+2\sqrt{\xi_{3}+\frac{1}{4}}, &; \alpha_{11}=2\sqrt{\xi_{1}}, &;
\alpha_{12}=\frac{1}{2}+\sqrt{\xi_{3}+\frac{1}{4}} \\
\alpha_{13}=-\sqrt{\xi_{1}}\end{array},
\end{equation}
We get

\begin{equation} \label{eq.76}
E_{n,l}=-\frac{2m}{4\hbar^{2}}\frac{\alpha^{2}}{(n+l+1)^{2}}
\end{equation}
and

\begin{equation} \label{eq.77}
F_{n,l}=C_{n,l}\zeta^{l+1}e^{-\frac{3}{2}}L_{n}^{(2l+1)}(\zeta)
\end{equation}
where $C_{n,l}$ the the normalization constant and $\zeta$ is
defined as

\begin{equation} \label{eq.78}
\zeta=\frac{2mze^{2}}{\hbar^{2}(n+l+1)}s.
\end{equation}
Solution of the angular part can be found in Refs.(38, 39).

\section{Conclusions}
The SE is solved for some certain potentials with a general
parametric approach. Solutions are obtained reducing the SE written
in the parametric form to a second order differential equation. In
the solution, the Nikiforov-Uvarov method is applied to get energy
eigenvalues and the corresponding wave functions. If the SE for a
given potential can be reduced to the NU differential form, this
procedure can be used to get the results easily.

\section{Acknowledgements}
This research was partially supported by the Scientific and
Technological Research Council of Turkey.

\end{document}